\documentclass[prb,twocolumn,amsmath,amssymb,superscriptaddress,preprintnumbers]{revtex4-1}
\usepackage[pdftex]{graphicx}
\usepackage[outdir=./]{epstopdf}
\usepackage{enumerate,ulem,bm}
\usepackage{here,color}
\DeclareMathOperator{\sech}{sech}

\newcommand{\bfr}{\boldsymbol{r}}
\newcommand{\bfk}{\boldsymbol{k}}

\begin{document}

\title{Localized charge in various configurations of magnetic domain wall in Weyl semimetal}
\author{Yasufumi Araki}
\affiliation{Institute for Materials Research, Tohoku University, Sendai 980-8577, Japan}
\affiliation{Frontier Research Institute for Interdisciplinary Sciences, Tohoku University, Sendai 980-8578, Japan}
\author{Akihide Yoshida}
\affiliation{Institute for Materials Research, Tohoku University, Sendai 980-8577, Japan}
\author{Kentaro Nomura}
\affiliation{Institute for Materials Research, Tohoku University, Sendai 980-8577, Japan}
\begin{abstract}
We numerically investigate the electronic properties of magnetic domain walls formed in a Weyl semimetal.
Electric charge distribution is computed from the electron wave functions,
by numerically diagonalizing the Hamiltonian under several types of domain walls.
We find a certain amount of electric charge localized around the domain wall,
depending on the texture of the domain wall.
This localized charge stems from the degeneracy of Landau states under the axial magnetic field,
which corresponds to the curl in the magnetic texture.
The localized charge enables one to drive the domain wall motion by applying an external electric field
without injecting an electric current,
which is distinct from the ordinary spin-transfer torque and is free from Joule heating.
\end{abstract}


\maketitle

\section{Introduction} 
\label{sec:introduction}
As a new class of three-dimensional (3D) topological materials,
Weyl semimetals (WSMs) have recently been attracting a great interest
\cite{Wan_2011,Burkov_2011,Zyuzin_2012}.
The electrons in a WSM are characterized by the conical band structure,
namely the ``Weyl cone'' structure,
around the band-touching points called ``Weyl points'' in momentum space.
The Weyl cone structure is realized by band inversion from strong atomic spin-orbit coupling,
which results in spin-momentum locking feature around the Weyl points.
While the band-touching points in Dirac semimetals are doubly degenerate due to time-reversal and spatial inversion symmetry \cite{Young_2012},
the Weyl points in WSMs are isolated each other in momentum space
by breaking either of these two symmetries.
The ``Fermi arc'' surface state,
which crosses the Fermi level by an open line in momentum space,
is one of the typical features of WSMs,
which connects the pair of Weyl points projected onto the surface Brillouin zone.
Using the angle-resolved photoemission spectroscopy (ARPES),
Weyl cones and surface Fermi arcs have recently been observed in WSMs with broken inversion symmetry,
such as in TaAs (Refs.~\onlinecite{Xu_S-Y_2015,Lv_B-Q_2015,Lv_B-Q_2015_2}) and NbAs (Ref.~\onlinecite{Xu_S-Y_2015_2}).

One of the important challenges in the context of WSM is to realize the WSM phase in a magnetic material.
Several classes of magnetic materials have been proposed as candidates for WSM,
such as ferromagnetic Heusler alloys (e.g. $\mathrm{Co_3 Sn_2 S_2}$; see Refs.~\onlinecite{Liu_E,Muechler,Xu_Q})
and chiral antiferromagnetic compounds (e.g. $\mathrm{Mn}_3 \mathrm{Sn}$; see Refs.~\onlinecite{Yang_2017,Ito,Kuroda}).
It was also suggested by a self-consistent calculation that the magnetic WSM phase can be realized
by introducing magnetic impurities to 3D topological insulators, such as $\mathrm{Bi_2 Se_3}$ (Ref.~\onlinecite{Kurebayashi_2014}).
In a magnetic WSM, time-reversal symmetry is broken and the Weyl points are split in momentum space,
corresponding to its magnetic order.
Requiring cubic symmetry to the system,
the low-energy Weyl Hamiltonian in the presence of a ferromagnetic order in the continuum limit
takes the simplified form,
\begin{align}
H &= v_{\rm F} \eta \bm{\sigma} \cdot \bm{p} -J \bm{\sigma} \cdot \bm{M}(\bm{r}),
\label{eq:cont-Hamiltonian}
\end{align}
where $\boldsymbol{\sigma}$ is the Pauli matrix for the electron spin.
$\boldsymbol{M}(\boldsymbol{r})$ denotes the background magnetization macroscopically averaged around position $\boldsymbol{r}$,
which is formed by the localized magnetic moments of the constituent magnetic elements.
The first term in Eq.~(\ref{eq:cont-Hamiltonian}) accounts for spin-momentum locking feature of the Weyl electrons,
with the Fermi velocity $v_\mathrm{F}$,
momentum operator $\boldsymbol{p} = -i\boldsymbol{\nabla}$,
and the chirality $\eta = \pm 1$ that labels each valley.
The exchange coupling between the localized magnetic moment and the electron spin is imprinted in the second term,
with the coupling constant $J$.

Rewriting Eq.~(\ref{eq:cont-Hamiltonian}),
we can regard the exchange coupling to the magnetic texture $\boldsymbol{M}(\bfr)$
as an ``axial vector potential'' $\bm{A}_5(\bm{r}) = (J/e v_{\rm F})\bm{M}(\bm{r})$,
which couples to each valley with the opposite sign $\eta = \pm$ as
\begin{align}
H = v_\mathrm{F} \eta \boldsymbol{\sigma} \cdot \left[\boldsymbol{p} - \eta e \boldsymbol{A}_5(\boldsymbol{r})\right]. \label{eq:cont-Hamiltonian-A5}
\end{align}
The analogy between $\boldsymbol{M}(\boldsymbol{r})$ and $\boldsymbol{A}_5(\boldsymbol{r})$ implies that
a magnetic texture, namely a nonuniform pattern in $\boldsymbol{M}(\boldsymbol{r})$,
can yield a significant effect on the Weyl electrons.
The curl in $\bm{A}_5(\bm{r})$ gives the \textit{axial magnetic field}
$\bm{B}_5(\bm{r}) = \boldsymbol{\nabla} \times \bm{A}_5(\bm{r})$.
Noncollinear magnetic textures, such as domain walls (DWs), skyrmions, helices, etc.,
accompany such an axial magnetic flux.
The axial electromagnetic fields alter the electronic states and transport
in a similar manner to the normal electromagnetic fields, except for the chirality dependence \cite{Liu_2013,Grushin,Pikulin}.
Based on the idea of the axial electromagnetic fields,
it was shown that the dynamics of magnetic textures in a WSM pumps a certain amount of electric charge \cite{Araki_pumping}.
It should be noted that such an analogy between magnetic textures and the effective electromagnetic fields
also applies to 2D Dirac electrons at topological insulator surfaces,
which accounts for the charge localization at a magnetic texture \cite{Nomura_2011}.

Among various types of topological magnetic textures,
magnetic DWs have long been studied intensively in many kinds of systems
to make use of them as carriers of information in spintronics devices (see Ref.~\onlinecite{Brataas} for review).
Information storage with an array of magnetic DWs, namely a magnetic racetrack memory,
was successfully tested by manipulating the DW motion in magnetic wires \cite{Parkin}.
While the DW motion can be directly induced by an external magnetic field,
a spin-polarized electric current can also drive the DW motion via the spin-transfer torque,
namely the angular momentum exchange between the electron spins and the local magnetic moments \cite{Berger,Saihi,Slonczewski,Ralph}.
In the presence of spin-orbit coupling, the torque induced by an electric current gets enhanced,
which is known as the spin-orbit torque \cite{Manchon_2008,Manchon_2009,Miron_2010,Miron_2011,Emori,Ryu_K-S,Shiomi_2014}.
However, energy loss from Joule heating is inevitable in those current-induced torques,
hence we need to find a more efficient method to control DWs electrically with smaller conduction current.

In the present paper,
we numerically investigate the electronic properties in a magnetic WSM in the presence of a DW.
The authors showed analytically in the previous paper \cite{Araki_DW}
that a certain amount of electric charge is localized at a DW in a magnetic WSM,
which implies that the DW can be manipulated by an external electric field.
However, this result relies on the simplified \textit{collinear} texture of DW
to make the Hamiltonian [Eq.~(\ref{eq:cont-Hamiltonian})] analytically solvable.
In this work, we focus on the electronic properties under more realistic DW textures,
by numerically diagonalizing the Hamiltonian on the hypothetical cubic lattice.
Looking at the band structure for each DW,
we observe the ``Fermi arc'' structure coming from the bound states at the DW.
Such bound states give rise to the localized charge around each DW,
which is directly calculated from the electron eigenfunctions.
We find that the amount and the distribution of the localized charge depends on the type of the DW,
which can be traced back to degeneracy of the Landau states
under the axial magnetic field corresponding to the DW.
On the other hand, the amount of the localized charge depends proportionally on the electron chemical potential
for any type of DW,
which makes it easy to tune and estimate the efficiency in electric manipulation of the DW.
We estimate the typical velocity of the DW driven by an electric field
from our calculation result.

This paper is organized as follows.
In Section \ref{sec:model},
we define the model Hamiltonian and the DW magnetic textures on a cubic lattice,
and calculate the band structure by numerically diagonalizing the Hamiltonian.
We focus on the behavior of zero modes and observe the ``Fermi arc'' structure arising at each DW.
In Section \ref{sec:excitation-energy},
we calculate the excitation energy of a single DW,
and compare the results among the three configurations of DWs.
In Section \ref{sec:localized-charge},
we observe the charge distribution in the presence of a single DW,
and focus on the electric charge localized around the DW.
We discuss the relationship between the localized charge (localized modes) and the axial magnetic field
corresponding to the magnetic texture.
Finally, in Section \ref{sec:conclusion},
we conclude our discussion and propose some future problems.
In this paper, we take $\hbar =1$ and restore it in the final numerical result.

\section{Model} 
\label{sec:model}
In this section, we define the model Weyl Hamiltonian on a hypothetical cubic lattice,
and introduce three realistic textures of magnetic DWs.
Using this model, we calculate the band structure numerically
and deal with the ``Fermi arc'' states arising from the DW.

\subsection{Lattice Hamiltonian} 
\label{sub:Hamiltonian}
In this paper, we treat Weyl electrons coupled with localized magnetic moments via the exchange interaction.
We here construct the model Hamiltonian on a hypothetical cubic lattice with the lattice spacing $a$,
\begin{align}
H &= H_{0} + H_{\rm exc} \\
H_{0} &= \sum_{i,j} c^{\dag}_{i} h_{ij} c_{j} + \mathrm{H.c.} \\
h_{ij} &= \frac{1}{2} \sum_{\mu} \left( -i t \alpha _\mu + r\beta \right) \delta _{i+\hat{\mu},j} + \frac{3r}{2}\beta\delta_{ij} \\
H_{\rm exc} &= -J \sum_{i} \bm{M}_{i} \cdot c^{\dag}_{i} \bm{\Sigma} c_{i}, 
\end{align}
where we have required cubic symmetry to the Weyl Hamiltonian for simplicity.
The Hamiltonian $H$ consists of the non-interacting Wilson-Dirac Hamiltonian $H_0$ (see Refs.~\onlinecite{Wilson,Qi_2008}) and the exchange interaction term $H_{\rm exc}$.
The 4-component fermionic operator
$c_i =(c_i^{R\uparrow}, c_i^{R\downarrow}, c_i^{L\uparrow}, c_i^{L\downarrow}) \ (c_i^\dag)$
annihilates (creates) an electron on the lattice site $i$,
with $R/L$ and $\uparrow/\downarrow$ denoting the electron chirality (valley) and spin, respectively.
The parameters $t$ and $r$ characterize the hopping in the Wilson-Dirac Hamiltonian,
defined between neighboring lattice sites $i$ and $j$, with $\mu=x,y,z$.
The $4 \times 4$ Dirac matrices $\bm{\alpha}$ and $\beta$ are defined by
\begin{align}
\bm{\alpha} &=
\begin{pmatrix}
\bm{\sigma} &0 \\
0 &-\bm{\sigma}
\end{pmatrix}, \quad
\beta =
\begin{pmatrix}
0 &I \\
I &0
\end{pmatrix},
\label{eq:alpha matrix}
\end{align}
with the Pauli matrices $\bm{\sigma}$ and the identity matrix $I$.
The exchange interaction $J$ couples the electron spin $c_i^\dag \bm{\Sigma} c_i$
with the localized magnetic moment $\bm{M}_i$,
where $\bm{\Sigma} = \bm{\sigma} \otimes I$ is the spin operator of the electrons.

This model can be diagonalized analytically under the uniform magnetization $\bm{M}_i = M_0 \bm{e}_x$.
Here the Hamiltonian under the periodic boundary condition can be rewritten by the Fourier transformation,
\begin{align}
H &= \sum_{\bm{k}} c^\dag_{\bm{k}} \left[ h_0(\bm{k}) + h_{\rm exc} \right] c_{\bm{k}} \\
h_0(\bm{k}) &= \sum_\mu \left( t\sin k_\mu a \right) \alpha_\mu + m(\bm{k}) \beta \\
m(\bm{k}) &= r \sum_\mu \left(1-\cos k_\mu a \right) \\
h_{\rm exc} &= JM_0\Sigma_x.
\end{align}
Thus the eigenvalues are given by 
\begin{align}
E_\tau (\bm{k}) &= \pm \sqrt{ \sum_{\mu = y,z}t^2\sin^2{k_\mu}a + \left[ \xi(\bm{k}) +\eta JM_0 \right]^2 } \\
\xi(\bm{k})^2 &= t^2\sin^2{k_x}a + m(\bm{k})^2
\end{align}
for each $\bfk$ in the Brillouin zone $\mathrm{BZ} = [-\pi/a,\pi/a)^3$,
with $\eta=\pm1$ the label for chirality $R/L$.
In case $r=t$, the valence and conduction bands touch at two Weyl points,
$\boldsymbol{K}_\pm = (\pm k_\Delta,0,0)$,
where
\begin{align}
k_\Delta = \frac{2}{a}\arcsin{ \left( \frac{JM_0}{2t} \right) }.
\label{eq:Weyl_point}
\end{align}
The separation of the Weyl points is determined by the magnitude and the direction of the background magnetization.
If the exchange energy is sufficiently small compared with the bandwidth, i.e. $JM_0/t\ll1$,
this form can be simplified as $k_\Delta \simeq JM_0/ta = JM_0/v_F$,
which is consistent with the continuum model at low energy shown by Eq.~(\ref{eq:cont-Hamiltonian}).
In the numerical calculations shown below,
we take all the physical quantities dimensionless:
we set the lattice spacing $a$ to unity,
and fix the parameters $t=2$, $r=2$ and $J|\bm{M}|=2$.

\begin{figure}[tb]
\begin{center}
\includegraphics[width=8cm]{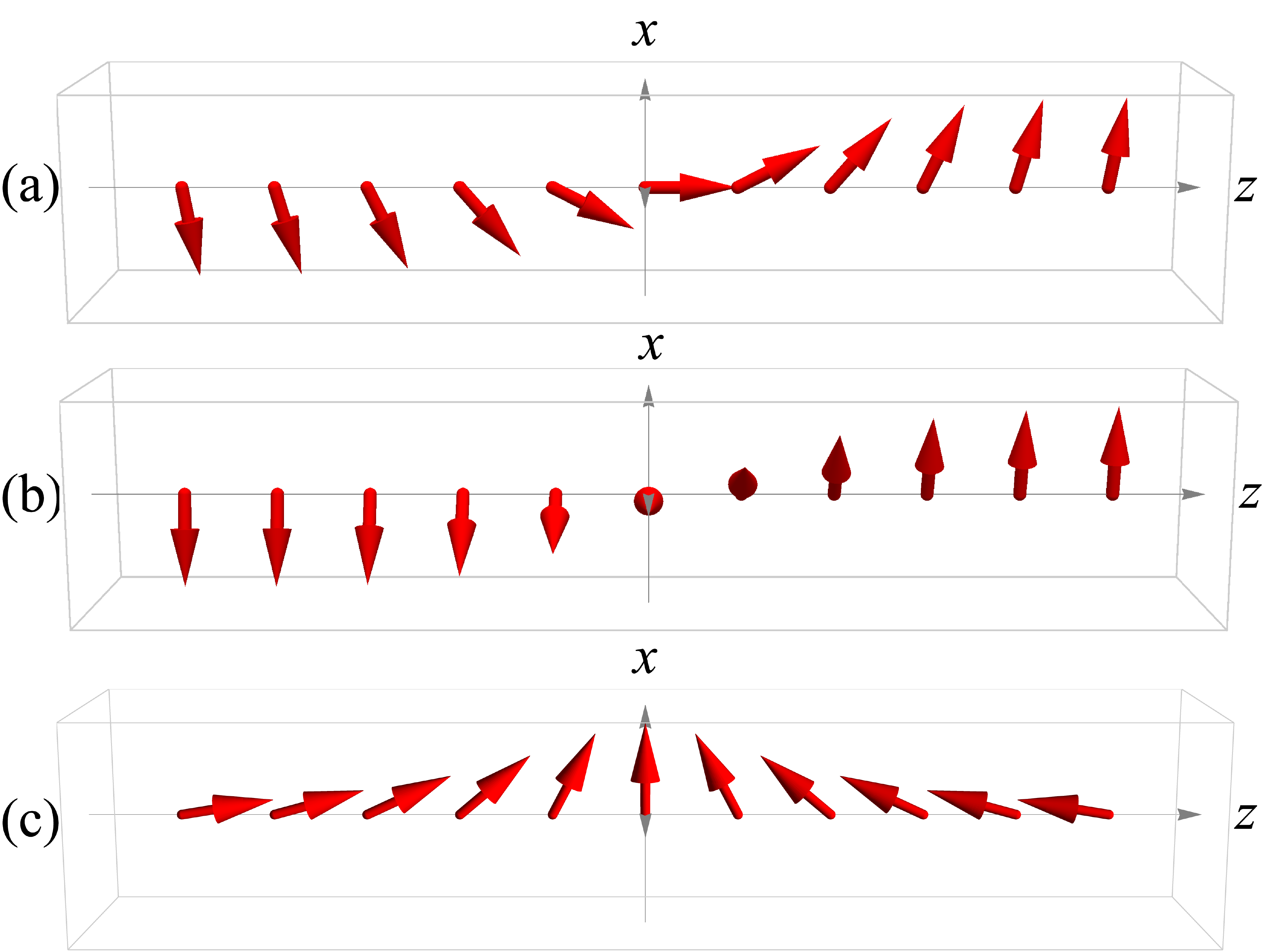}
\caption{Schematic pictures of DW configurations employed in this paper.
Red arrows denote the directions of localized magnetic moments.
DWs are set up around the $xy$-plane $(z=0)$.
(a)(b)(c) correspond to the \textit{coplanar} (Ne\'{e}l), \textit{spiral} (Bloch),
and \textit{head-to-head} DWs, respectively.}
\label{fig:DW}
\end{center}
\end{figure}

\subsection{Band structure under domain wall} 
\label{sub:Fermi arc under DW}
Now we are interested in the case where the local magnetization $\boldsymbol{M}_i$ forms a certain texture.
Here we introduce three types of DW strctures,
and calculate the eigenvalues and eigenstates by numerically diagonalizing the lattice Hamiltonian.
Taking the DW parallel to the $xy$-plane,
$x$- and $y$-directions can be diagonalized analytically by the Fourier transformation,
while $z$-direction is no longer translationally invariant and hence it is treated numerically.
Thus, we treat the system as a series of slabs stacked in $z$-direction.
We introduce three types of configurations of the DW.
There are two ways to twist the magnetization from $\bm{M}=-M_0\bm{e}_x$ on the one end to $\bm{M}=M_0\bm{e}_x$ on the other, as shown in Fig.~\ref{fig:DW} (a) and (b).
In case (a), the magnetization points perpendicular to the DW at the center of the DW,
which is called ``Ne\'{e}l DW'' or ``\textit{coplanar} DW''.
In case (b), the magnetization is always in the DW plane,
which is called ``Bloch DW'' or ``\textit{spiral} DW''.
On the other hand, twisting pattern of the magnetization
from $\bm{M}=M_0\bm{e}_z$ to $\bm{M}=-M_0\bm{e}_z$ can be uniquely constructed due to cubic symmetry,
as shown in Fig.~\ref{fig:DW} (c),
which we refer to as the ``\textit{head-to-head} DW''.

We first focus on the band structure under these DWs.
We here set the periodic boundary condition for all directions, to exclude the effect from the surface.
The size of the system in $z$-direction is given by $L_z=N_z a$,
where the number of slabs $N_z$ is fixed to 32 here.
The $x$- and $y$- directions are Fourier transformed due to the translational invariance,
with the wave vector $(k_x, k_y)$.
Therefore, the eigenstates and the eigenvalues can be labeled as $\psi_{k_x,k_y,n}(z)$ and $\epsilon_{k_x,k_y,n}$,
where $n$ is the label for energy level.
The wave function $\psi_{k_x,k_y,n}(z)$ consists of four components in the spin and pseudospin (chirality) spaces.

The \textit{coplanar} and \textit{spiral} DWs under the periodic boundary condition are defined by
\begin{align}
\bm{M}_{\rm c}(z) &= M_0
\left( \cos{2\pi \frac{z}{L_z}}, \ 0, \ \sin{2\pi \frac{z}{L_z}} \right) \\
\bm{M}_{\rm s}(z) &= M_0 \left( \cos{2\pi \frac{z}{L_z}}, \ \sin{2\pi \frac{z}{L_z}}, \ 0 \right),
\end{align}
respectively,
which consist of a pair of opposite DWs per each period,
namely a magnetic spiral.
Since the \textit{coplanar} and the \textit{head-to-head} configurations are equivalent
by a translation by $L_z/4$ in the $z$-direction in this case,
we do not take into account the band structure under the \textit{head-to-head} configuration explicitly.	
We also take the uniform magnetization as a reference,
where all the localized magnetic moments are pointing in $x$-direction $(\bm{M}=M_0\bm{e}_x)$.

\begin{figure}[tb]
\begin{center}
\includegraphics[width=8cm]{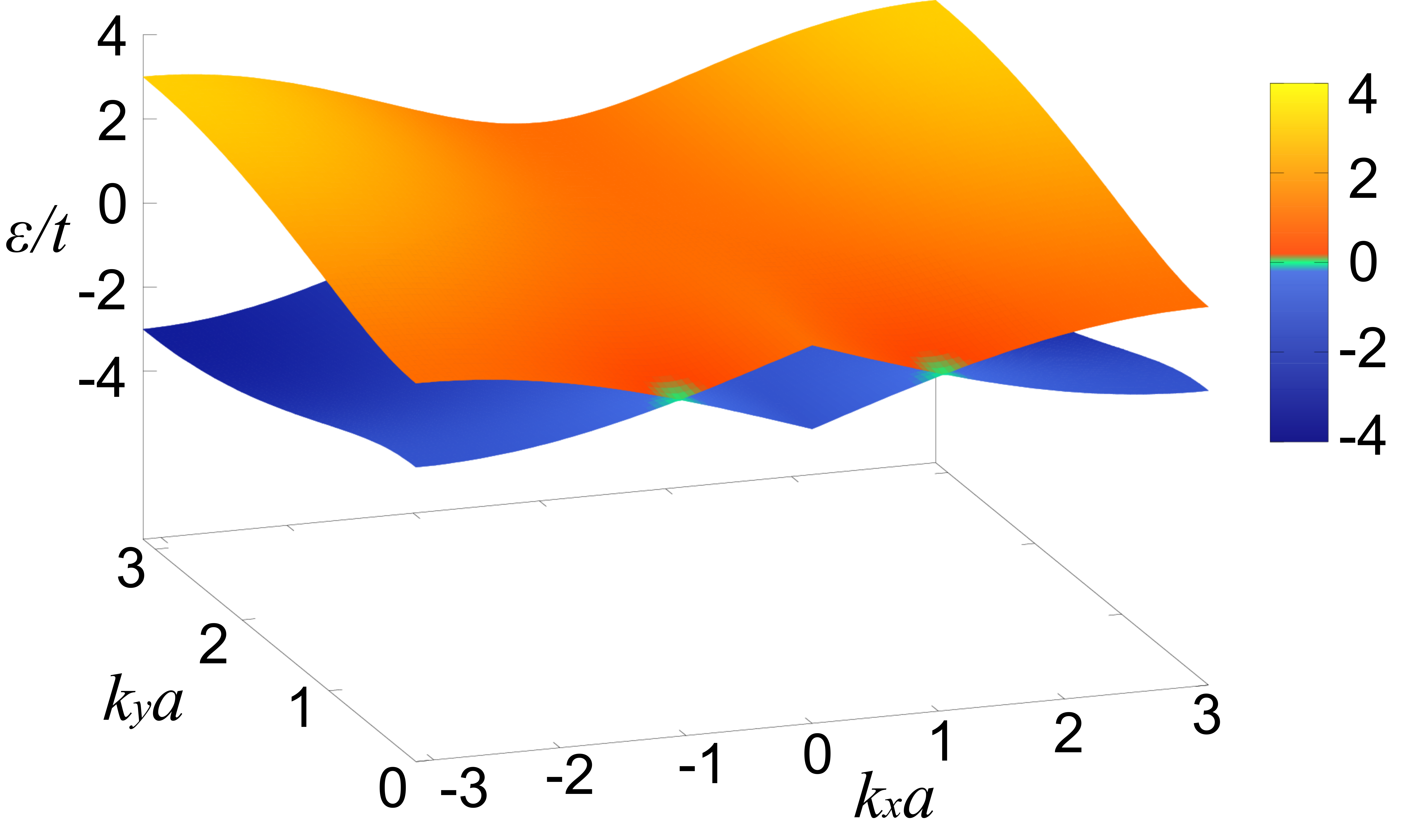}
\caption{Band structure under the uniform magnetization $\bm{M}=M_0\bm{e}_x$,
picking up the two bands crossing at zero energy.
The region in the vicinity of zero energy is shown in green.
Only half of the Brillouin zone $(k_y>0)$ is shown to clarify the crossing points of the Weyl cones.}
\label{fig:FA_uniform}
\end{center}
\end{figure}

\begin{figure}[tb]
\begin{center}
\includegraphics[width=8cm]{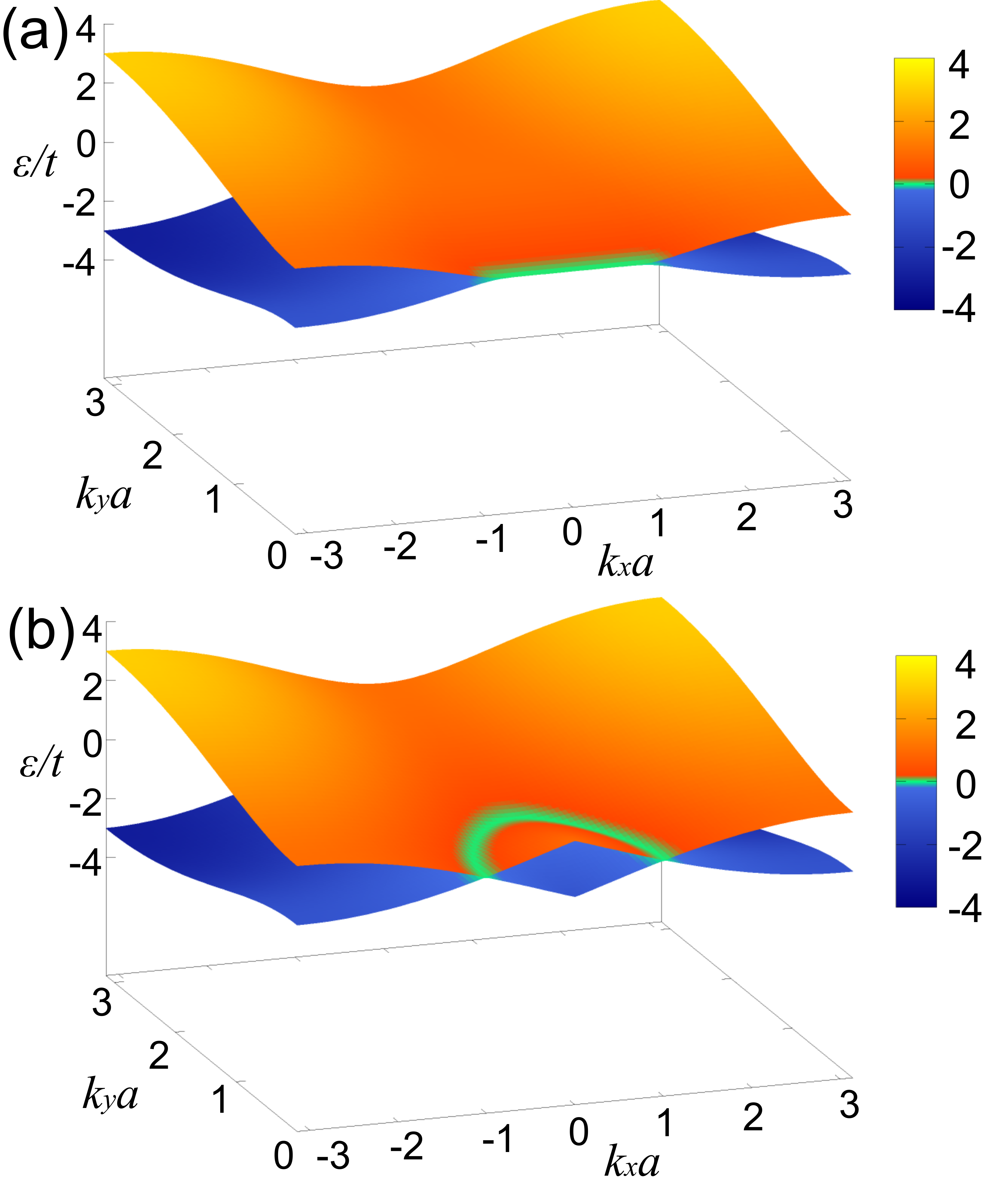}
\caption{Band structures under the \textit{coplanar} and \textit{spiral} magnetization textures,
with two bands picked up as in Fig.~\ref{fig:FA_uniform}.
(a) The electronic spectrum under the \textit{coplanar} DW shows the straight ``Fermi arc'' between two Weyl points.
(b) The Fermi arc for the \textit{spiral} DW shows the circular structure,
following the trajectory of the Weyl points.}
\label{fig:FA_DW}
\end{center}
\end{figure}

By numerically diagonalizing the lattice Hamiltonian,
we straightforwardly obtain the band structure
as a function of the in-plane wave vector $(k_x, k_y)$ for each DW configuration.
We here extract two bands crossing at zero energy,
which may show the most typical behavior among all bands.
Due to particle-hole symmetry,
there emerge electron (orange) and hole (blue) bands touching at zero energy,
as shown in Figs.~\ref{fig:FA_uniform} and \ref{fig:FA_DW}.
Under the uniform magnetization $\boldsymbol{M} = M_0 \boldsymbol{e}_x$,
time-reversal symmetry is broken and the two Weyl points are split in $k_x$-direction, as estimated in Eq.~(\ref{eq:Weyl_point}) (see Fig.~\ref{fig:FA_uniform}).

On the other hand, in the presence of a DW,
the electron and hole bands touch by a line at zero energy, namely the ``Fermi arc'',
instead of the isolated Weyl points.
As can be seen in Fig.~\ref{fig:FA_DW}, the \textit{coplanar} DW gives a linear Fermi arc,
while the \textit{spiral} DW yields a circular arc (closed loop).
The ``Fermi arc'' observed here can be regarded as the trajectory of the Weyl points;
Equation (\ref{eq:cont-Hamiltonian-A5}) states that,
under a (locally) uniform magnetization $\boldsymbol{M}$,
the location of the Weyl point for each valley $\eta$ is determined straightforwardly by $\boldsymbol{M}$,
as $\boldsymbol{K}_\eta = \eta \boldsymbol{A}_5 = \eta J\bm{M}/v_{\rm F}$.
As long as the spatial variation of $\bm{M}(z)$ is sufficiently slow, the positions of the Weyl points in the $(k_x,k_y)$-plane can well be regarded as a function of $z$.
By following the magnetization from $z=0$ to $L_z$, the Weyl points draw either linear or circular trajectories,
depending on the type of the DW.

Since the Fermi arc found here emerges inside the DW,
it cannot be measured directly, e.g. by ARPES.
However, it alters the electron energy and the charge distribution of the system,
which we shall discuss in the following sections.

\section{Excitation energy of domain wall} 
\label{sec:excitation-energy}
In this section, we estimate the excitation energy of a DW,
to find out which DW configuration is the most stable.
We first calculate the excitation energy numerically
by comparing the total electron energy with that under a uniform magnetization.
We also compare our results with the free energy that was analytically derived in the continuum limit in the previous work \cite{Araki_corr}.

\subsection{Numerical calculation}
Here we investigate the effect of a single DW fixed around $z=0$, with $z \in [-L_z/2, L_z/2]$.
The sizes of the system in $x$- and $y$-direction are given by $L_x=N_x a$ and $L_y=N_y a$,
and we take here $N_x = N_y \simeq 1000$;
note that $N_x$ and $N_y$ only determine the mesh in $(k_x, k_y)$-plane and have no physical importance.
We employ the open boundary condition at $z=\pm L_z/2$ (i.e. exposed to vacuum),
while $x$- and $y$-directions are set periodic.
The DW configuration for each type is defined by
\begin{align}
\bm{M}_{\rm c}(z) &= M_0 \left(\tanh \frac{z}{W}, \ 0, \ \sech \frac{z}{W} \right)  \label{eq:DW-coplanar} \\
\bm{M}_{\rm s}(z) &= M_0 \left(\tanh \frac{z}{W}, \ \sech \frac{z}{W}, \ 0 \right) \label{eq:DW-spiral} \\
\bm{M}_{\rm h}(z) &= M_0 \left(\sech \frac{z}{W}, \ 0, \ -\tanh \frac{z}{W} \right), \label{eq:DW-head-to-head}
\end{align}
with $W$ the width of the DW.
The subscripts c, s, and h represent the \textit{coplanar}, \textit{spiral}, and \textit{head-to-head} DWs, respectively,
which we shall use in the results below.

\begin{figure}[tb]
\begin{center}
\includegraphics[width=8cm]{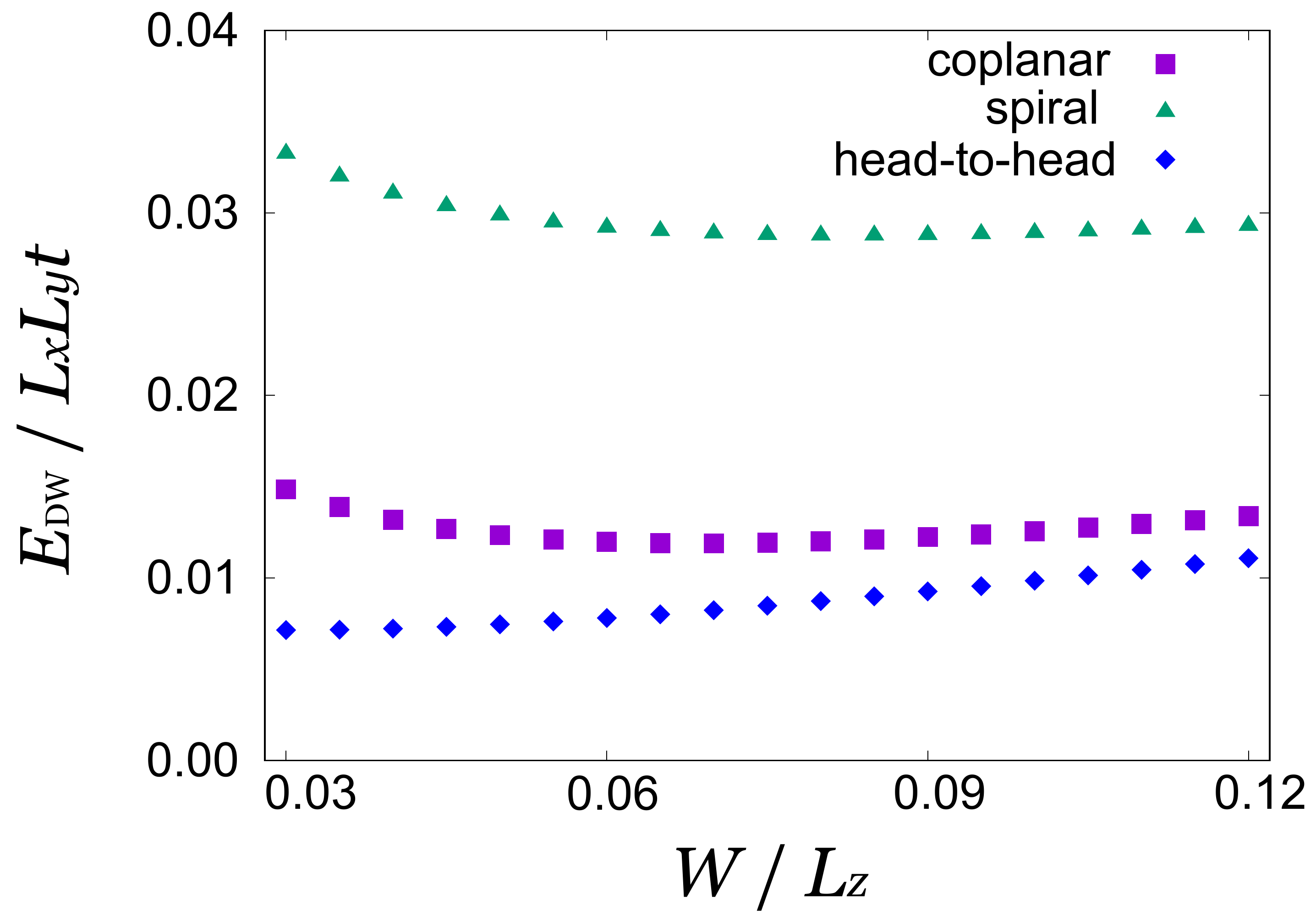}
\caption{The excitation energy of DW at $\mu=0$ as a function of the DW width $W$,
which is obtained from Eq.~(\ref{eq:excitation-energy}).
The excitation energy of the \textit{spiral} DW is the largest among all the three DW textures.
}
\label{fig:GE}
\end{center}
\end{figure}

The excitation energy of a DW is given by evaluating the energy of the electrons in the occupied states.
Taking the electron chemical potential $\mu$ fixed to charge neutrality,
the energy per unit area to excite a DW from the uniform magnetization is defined by
\begin{align}
E_{\rm DW} = \frac{1}{L_x L_y}\sum_{\epsilon\leq0}\epsilon_{k_x,k_y,n}^{\rm (DW)} - \frac{1}{L_x L_y}\sum_{\epsilon\leq0}\epsilon_{k_x,k_y,n}^{\rm (uniform)}, \label{eq:excitation-energy}
\end{align}
where $\epsilon_{k_x, k_y, n}^\mathrm{(DW)}$ is the single-particle energy for each eigenstate under the DW,
while $\epsilon_{k_x, k_y, n}^\mathrm{(uniform)}$ is the energy under the uniform magnetization.
The reference uniform magnetization is taken
$\bm{M}=M_0\bm{e}_x$ for the \textit{coplanar} and \textit{spiral} DWs,
and $\bm{M}=M_0\bm{e}_z$ for the \textit{head-to-head} DW.
Figure \ref{fig:GE} shows $E_\mathrm{DW}$ for each DW configuration as a function of the DW size $W$;
we can see that the \textit{spiral} DW requires the largest excitation energy among the three types,
while the \textit{head-to-head} DW the smallest.
It can be qualitatively understood in terms of the number of zero modes.
We have seen that there evolves a new ``Fermi arc'' state connecting the two Weyl points
by introducing a DW to the magnetic texture,
which means that a DW excites a bulk electron on the Weyl cone in the Fermi sea
onto the Fermi arc at zero energy.
Since the Fermi arc under the \textit{spiral} DW is longer than that under the \textit{coplanar} DW,
we can estimate that the number of zero modes 
under the \textit{spiral} DW is larger than that under the \textit{coplanar} DW.
Thus the DW Fermi arc accounts for the difference of the excitation energy depending on the DW configuration.

\subsection{Domain wall energy in continuum limit}
If the DW size $W$ is large enough,
we can estimate the DW energy analytically in terms of the gradient expansion as well.
Starting from the continuum Hamiltonian [Eq.~(\ref{eq:cont-Hamiltonian})]
and integrating out the fermionic degrees of freedom,
the free energy functional $F[\bm{M}]$ for the magnetic texture $\boldsymbol{M}(\bfr)$
was obtained in Ref.~\onlinecite{Araki_corr} as
\begin{align}
F[\bm{M}] &= F_0[\bm{M}] + F_1[\bm{M}] \\
F_1[\bm{M}] &= \int d^3\bm{r} \left( J_{\rm i} \left[\boldsymbol{\nabla}\boldsymbol{M}\right]^2 + J_{\rm a} \left[\boldsymbol{\nabla}\times\boldsymbol{M}\right]^2 \right), \label{eq:F1}
\end{align}
with shorthand notations $\left[\boldsymbol{\nabla}\boldsymbol{M}\right]^2=\sum_{i,j}\left(\partial_iM_j\right)^2$ and 
$\left[\boldsymbol{\nabla}\times\boldsymbol{M}\right]^2=\sum_i\left(\boldsymbol{\nabla}\times\boldsymbol{M}\right)_i^2$.
The coefficients $J_\mathrm{i}$ and $J_\mathrm{a}$ are given as
\begin{align}
J_\mathrm{i} = \frac{J^2}{48\pi^2 v_\mathrm{F}^3} \frac{1}{10}, \quad 
J_\mathrm{a} = \frac{J^2}{48\pi^2 v_\mathrm{F}^3} \left[\ln \frac{k_\mathrm{C}}{k_\mathrm{F}} -\frac{11}{15}\right],
\end{align}
where $k_\mathrm{F} = \mu/v_\mathrm{F}$ is the Fermi momentum
and $k_\mathrm{C}$ is the momentum cutoff that characterizes the size of the Brillouin zone.

$F_0[\bm{M}]$ is the homogeneous part of the free energy,
which is determined only by the magnitude of the local magnetization (i.e. $|\boldsymbol{M}(\boldsymbol{r})|^2$)
and makes no difference among the DW configurations employed here.
On the other hand, the inhomogeneous part $F_1[\bm{M}]$ depends on the spatial texture,
and is minimized under the uniform magnetization.
This part consists of two terms:
the first term with the coefficient $J_\mathrm{i}$ comes from the Fermi surface.
This term has the same form as the conventional Heisenberg-like RKKY interaction,
namely the effective interaction between localized spins mediated by the spin of conduction electrons.
The second term with $J_\mathrm{a}$ originates from the interband transition process between the electron and hole bands
induced by the scattering by a localized magnetic moment, which is known as the Van Vleck mechanism.
Spin mixing by spin-orbit coupling is essential in the Van Vleck mechanism,
hence strong spin-momentum locking around the Weyl points in this system largely enhances this term,
which is captured as the logarithmic divergence in $J_\mathrm{a}$ for $k_\mathrm{F} \rightarrow 0$.
Therefore, in the vicinity of the Weyl points (i.e. $\mu \sim 0$),
$J_\mathrm{a}$-term is dominant and $J_\mathrm{i}$-term can be safely neglected.

Using Eq.~(\ref{eq:F1}) and neglecting $J_\mathrm{i}$-term,
the free energy per unit area $f_1 = F_1/A$ for each DW configuration is given by
\begin{align}
f_1[\boldsymbol{M}_\mathrm{c}]= \frac{4J_{\rm a}}{3W}, \
f_1[\boldsymbol{M}_\mathrm{s}]= \frac{2J_{\rm a}}{W}, \
f_1[\boldsymbol{M}_\mathrm{h}]= \frac{2J_{\rm a}}{3W},
\end{align}
with $A$ the area of the DW.
We can see that the \textit{spiral} DW requires the largest excitation energy
and the \textit{head-to-head} the smallest in continuum limit,
which is a complementary result with our numerical results on the lattice.
The quantitative discrepancy comes from
the sharp spatial variation in the magnetic texture around the center of the DW,
which makes the gradient expansion less reliable.

\section{Localized charge}
\label{sec:localized-charge}
In this section, we focus on the electric charge distribution altered by the DW,
using the eigenstate wavefunctions obtained by the numerical diagonalization.
In a WSM with a uniform magnetization,
all the eigenstates are composed of plane waves
and consequently the charge distribution is uniform in the bulk.
On the other hand,
a magnetic DW in a WSM hosts a linearly dispersed Fermi arc mode,
which may contribute to the localized charge.
Here we numerically calculate the deviation of the charge distribution under the DWs
from that in the uniform system.
We show that a certain amount of charge gets localized around the DW,
and calculate the amount of the localized charge for each DW configuration.
We shall employ the same DW configuration as in the previous section for our numerical calculation.

\subsection{Charge distribution} 
\label{sub:Electric charge distribution}
\begin{figure}[tb]
\begin{center}
\includegraphics[width=8cm]{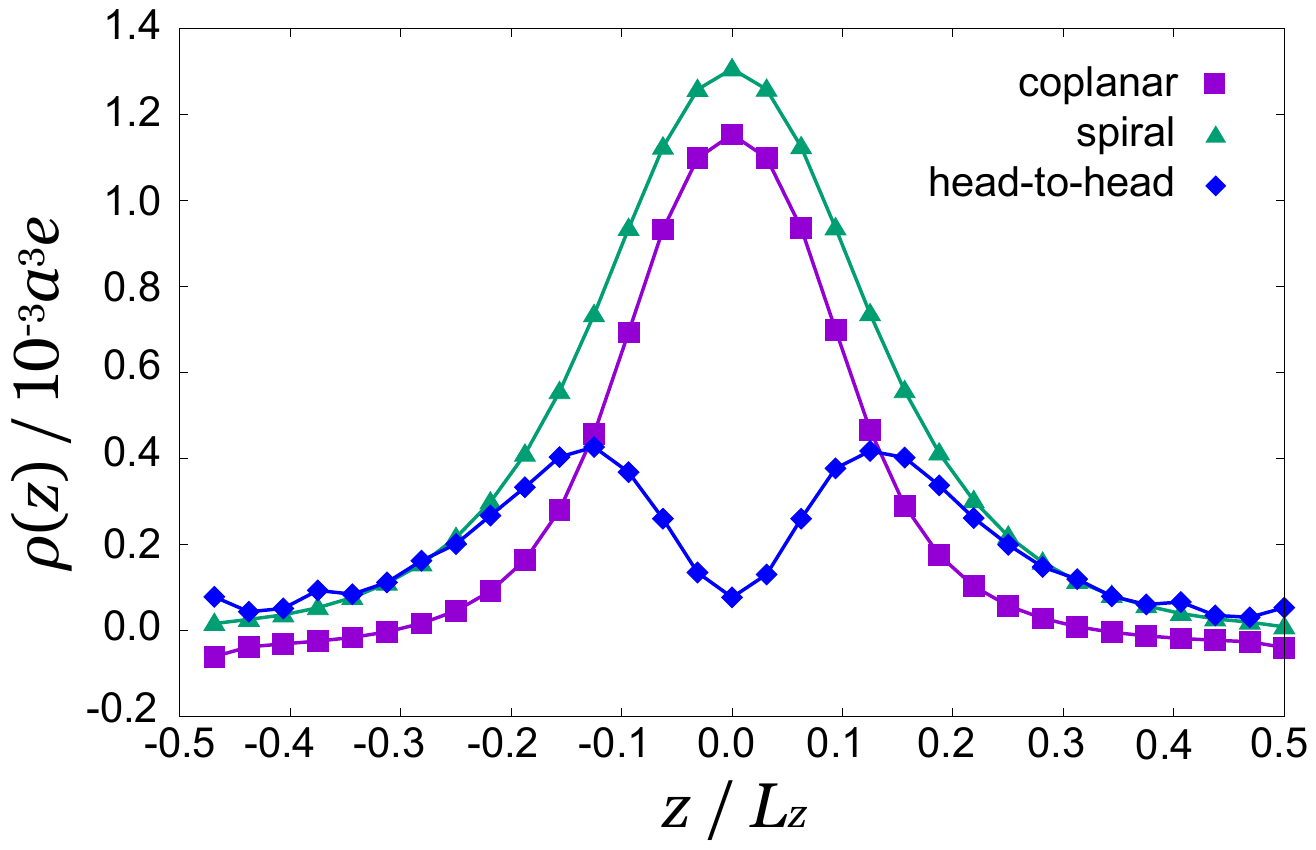}
\caption{The charge distribution calculated by Eq.~(\ref{eq:charge-distribution}),
with $\mu /t=0.1$ and $W/L_z=0.1$.
The \textit{coplanar} and the \textit{spiral} DWs give a single large peak,
while the \textit{head-to-head} DW gives two small peaks.}
\label{fig:CD}
\end{center}
\end{figure}

Taking the summation over the wave function distributions over all the occupied eigenstates, 
we can estimate the charge distribution induced by the DW as
\begin{align}
\delta\rho(z) = \frac{e}{L_x L_y a}\sum_{k_x,k_y}\sum_{\epsilon\leq\mu} \left[ |\psi_{k_x,k_y,n}^{\rm (DW)}(z)|^2 - |\psi_{k_x,k_y,n}^{\rm (uniform)}(z)|^2 \right], \label{eq:charge-distribution}
\end{align}
with $e$ an elementary charge, for each DW configuration.
Here $\psi_{k_x,k_y,n}^\mathrm{(DW)}(z)$ and $\psi_{k_x,k_y,n}^\mathrm{(uniform)}(z)$ denote
the eigenfunction under the DW texture and that under the uniform magnetization, respectively.
Figure \ref{fig:CD} shows our calculation result,
for the chemical potential $\mu = 0.1 t$ and the DW size $W = 0.1 L_z$.
We can see here that charge is accumulated around the DW for all the three DW configurations.
We should note that the \textit{coplanar} and the \textit{spiral} DWs give a single peak at the center of the DW,
while the \textit{head-to-head} type gives two peaks separated by a small cusp at the center.

The charge localization found here can be traced back to the Landau quantization under the \textit{axial} magnetic field.
As we have seen in Eq.~(\ref{eq:cont-Hamiltonian-A5}),
the local magnetization $\boldsymbol{M}(\bfr)$ couples to the electron spin
as an axial gauge field $\boldsymbol{A}_5(\bfr)=(J/e v_F)\boldsymbol{M}(\bfr)$ at low energy.
The magnetic DW texture bears a vorticity inside the DW, which corresponds to the axial magnetic field,
\begin{align}
\boldsymbol{B}_5(\bm{r}) = \boldsymbol{\nabla} \times \boldsymbol{A}_5(\bfr).
\end{align}
Using the formulations in Eqs.~(\ref{eq:DW-coplanar})-(\ref{eq:DW-head-to-head}),
the axial magnetic field distribution for each DW configuration is given by
\begin{align}
\boldsymbol{B}_{5 \mathrm{c}}(z) &= \frac{JM_0}{e v_{\rm F}} \left( 0,\ \sech^2 \frac{z}{W},\ 0 \right) \\
\boldsymbol{B}_{5 \mathrm{s}}(z) &= \frac{JM_0}{e v_{\rm F}} \left(\tanh \frac{z}{W} \sech \frac{z}{W},\ \sech^2 \frac{z}{W},\ 0 \right) \\
\boldsymbol{B}_{5 \mathrm{h}}(z) &= \frac{JM_0}{e v_{\rm F}} \left(0,\ -\tanh \frac{z}{W} \sech \frac{z}{W},\ 0\right).
\end{align}
It should be noted that the direction of the axial magnetic field,
i.e. $\hat{\boldsymbol{B}}_5(z) \equiv \boldsymbol{B}_5(z)/|\boldsymbol{B}_5(z)|$,
is unidirectional ($y$-direction) for the \textit{coplanar} DW,
while it spatially varies for the \textit{spiral} and \textit{head-to-head} cases.
$\boldsymbol{B}_{5 \mathrm{h}}(z)$ suddenly flips its direction at the DW center,
while $\boldsymbol{B}_{5 \mathrm{s}}(z)$ gradually changes its direction inside the DW.

\begin{figure}[tb]
\begin{center}
\includegraphics[width=7cm]{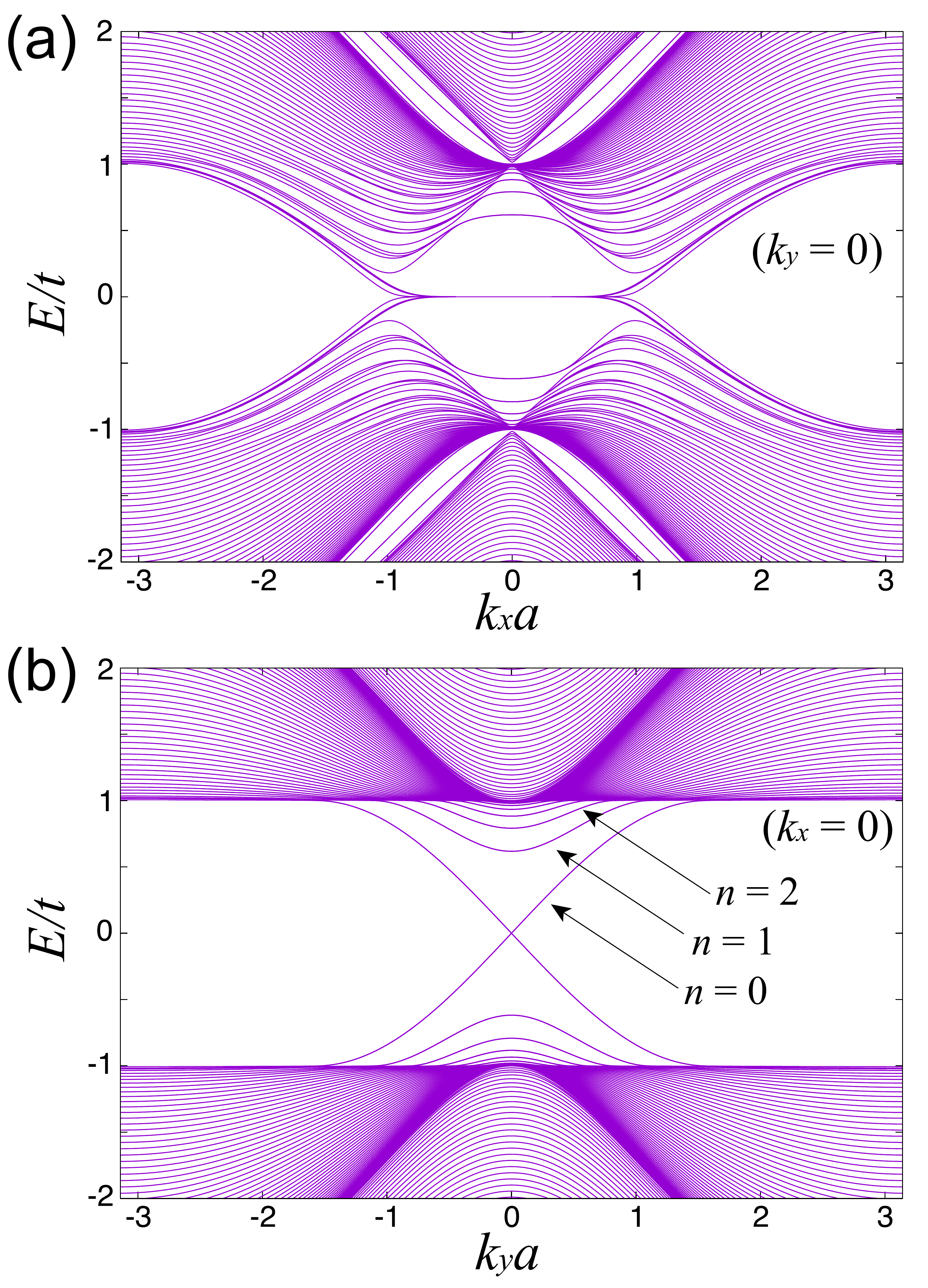}
\caption{The band structure in the presence of two opposite \textit{coplanar} DWs
under the periodic boundary condition,
with (a) $k_y =0$ and (b) $k_x =0$.
The bands show the Landau-like spectrum around zero energy,
with the zeroth LL $(n=0)$ unidirectionally dispersed in $k_y$-direction.}
\label{fig:DW_bands}
\end{center}
\end{figure}

The axial magnetic field induces the Landau quantization in the electronic spectrum,
just like a normal (realistic) magnetic field \cite{Liu_2013}.
In the presence of a uniform axial magnetic field $\boldsymbol{B}_5 = B_5 \boldsymbol{e}_y$, for instance,
the Landau level (LL) spectrum for the $n$-th LL $(n =0, \pm 1 ,\pm 2 , \cdots)$ is given by
\begin{align}
E_{n \neq 0}(k_y) &= v_\mathrm{F} \mathrm{sgn}(n)\sqrt{2|n|e B_5 + k_y^2} \\
E_{n = 0}(k_y) &= v_\mathrm{F} k_y,
\end{align}
where the zeroth LL is unidirectionally dispersed along the axial magnetic field \cite{Liu_2013,K-Y_Yang}.
Figure \ref{fig:DW_bands} shows the band structure in the presence of \textit{coplanar} DWs
calculated by the numerical diagonalization \cite{DW-note};
it shows the Landau-like spectrum around zero energy,
which comes from the axial magnetic field arising from the DW texture.

\begin{figure}[tb]
\begin{center}
\includegraphics[width=8cm]{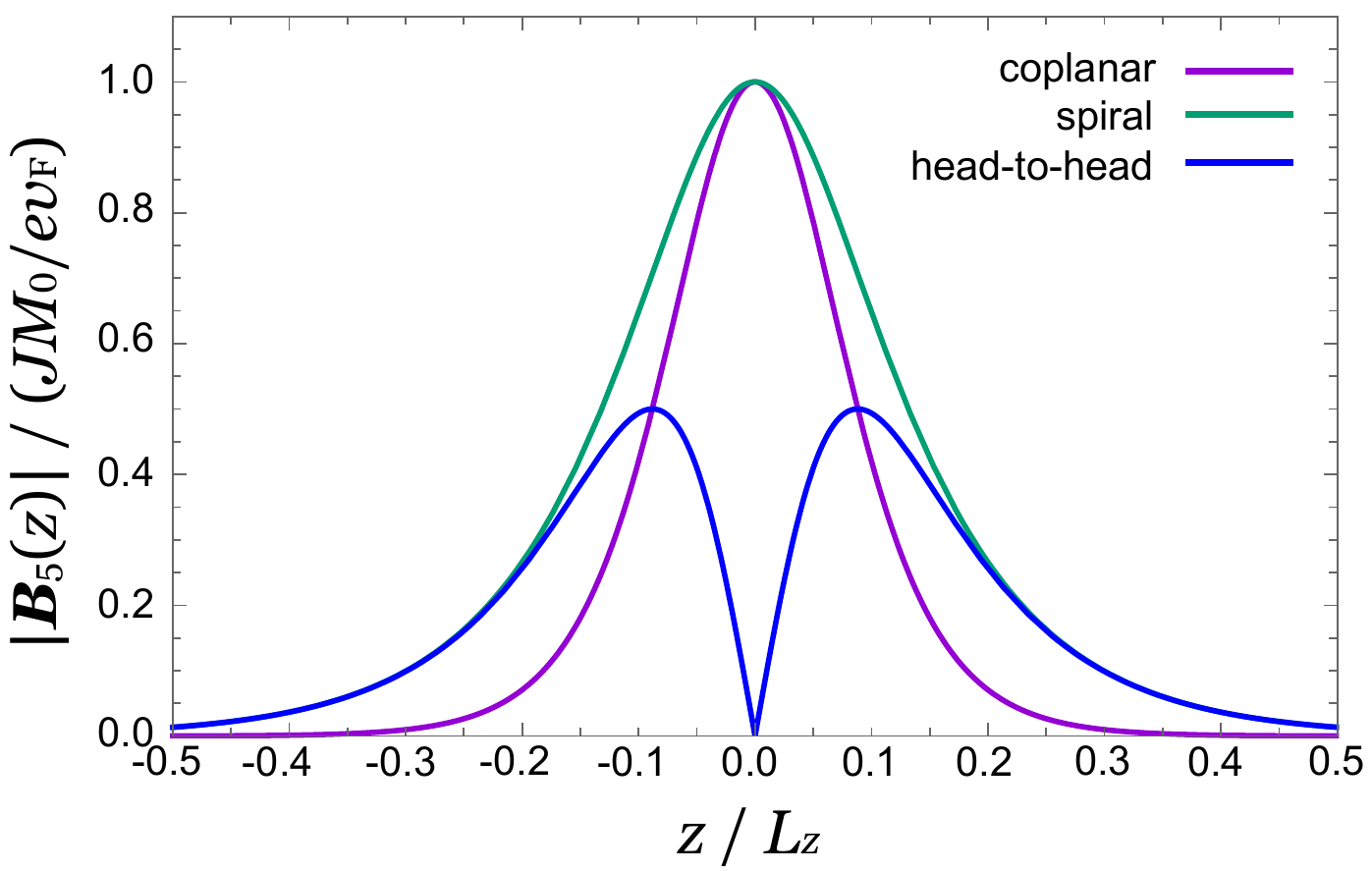}
\caption{The strength of the axial magnetic field $\boldsymbol{B}_5$ for each DW configuration.
$|\boldsymbol{B}_5|$ for the \textit{head-to-head} DW has two peaks and vanishes at $z=0$,
which accounts for the cusp structure in the charge distribution seen in Fig.~\ref{fig:CD}.}
\label{fig:AMF}
\end{center}
\end{figure}

Here we should note that the Landau states localize at the magnetic flux,
with the degeneracy $\rho = B_5/\phi_0$ per unit area,
where $\phi_0=e/2\pi$ is the flux quantum.
This implies that, in the presence of a spatial variation in $\boldsymbol{B}_5$,
the charge density becomes larger in the region where the amplitude of $\boldsymbol{B}_5$ is strong.
Figure \ref{fig:AMF} shows the spatial distribution of $|\boldsymbol{B}_5(z)|$ for each DW texture,
all of which show strong $|\boldsymbol{B}_5|$ around the center of the DW.
Comparing the charge distribution in Fig.~\ref{fig:CD} and the axial magnetic field distribution in Fig.~\ref{fig:AMF},
we can regard the charge localization as the result of the Landau quantization and degeneracy by the axial magnetic field at the DW.
For instance, the \textit{spiral} DW gives the strongest axial magnetic field among the three DW configurations, 
which leads to the largest localized charge, as seen in the calculation.
The two-peak structure in the charge distribution in the presence of the \textit{head-to-head} DW
can also be traced back to the strength of the corresponding axial magnetic field.
Since the wave function of each Landau state has a tail at a scale of the cyclotron radius,
the charge density at $z=0$ becomes finite even though the axial magnetic field reaches zero there.

If the Fermi level is fixed to zero energy,
the contribution from the occupied LLs is negligible;
while the charge density coming from the bulk states is proportional to the volume of the Brillouin zone,
namely $k_\mathrm{C}^3$,
that from the Landau states is proportional to the length of the Brillouin zone, namely $\rho k_\mathrm{C}$.
Therefore, at charge neutrality,
the charge distribution becomes completely flat in the continuum limit (i.e. $k_\mathrm{C} \rightarrow \infty$),
except for the surface of the sample.
On the other hand,
when the Fermi level is set slightly above or below zero energy,
the deviation of the charge distribution from charge neutrality
mainly comes from the zeroth LL;
the density of states in the zeroth LL is constant around $\mu=0$,
while that for Weyl cones is quadratic in $\mu$.
The relation between the localized charge and the chemical potential
shall be discussed further in the next subsection.

\subsection{Total localized charge} 
\label{sub:Total electric charge}
\begin{figure}[tb]
\begin{center}
\includegraphics[width=8cm]{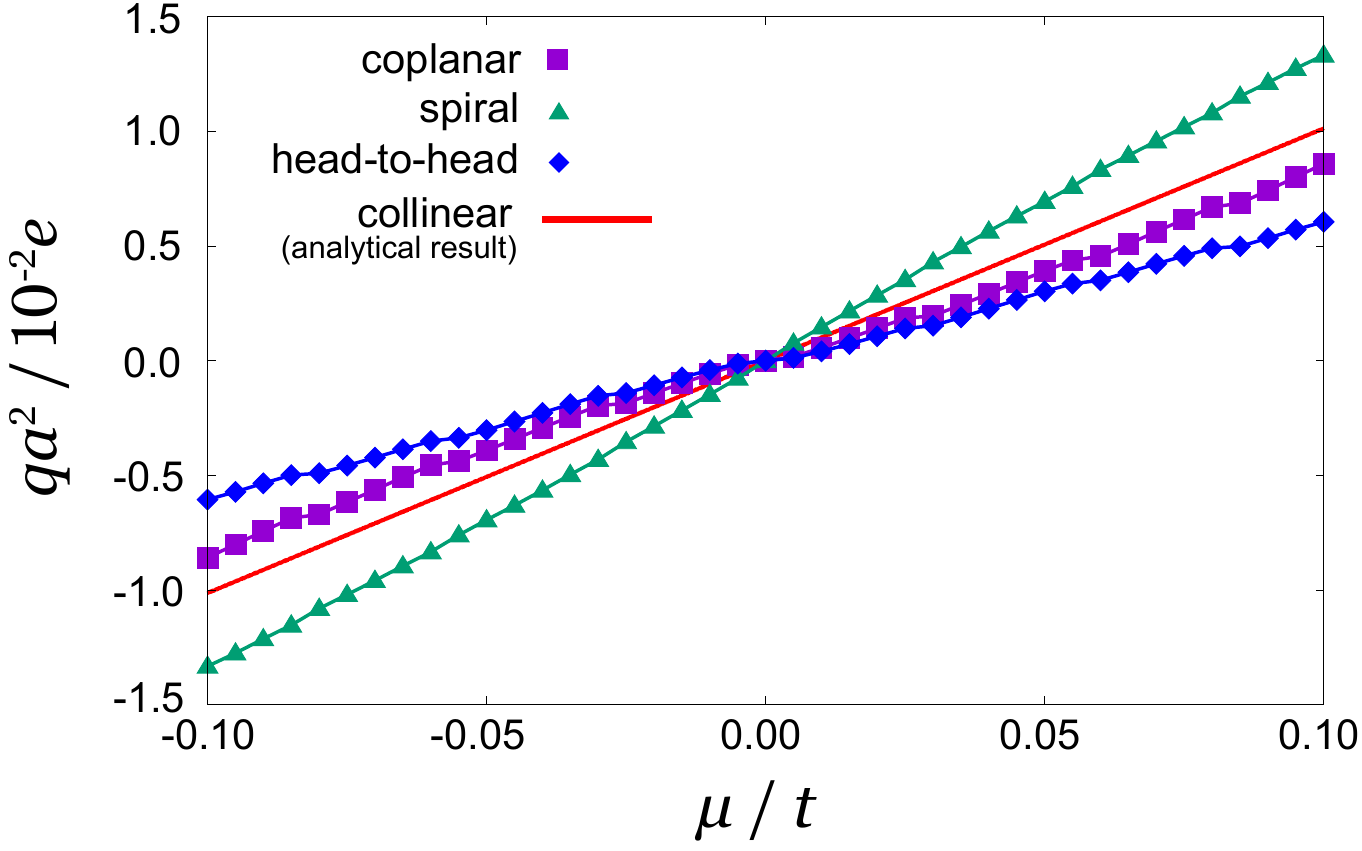}
\caption{The net localized charge $q$ as a function of the chemical potential $\mu$ for each DW,
with the DW size $W = 0.1 L_z$.
The net localized charge at the \textit{collinear} DW,
which was analytically obtained in Ref.~\onlinecite{Araki_DW} [Eq.~(\ref{eq:charge-collinear})],
is shown by the red solid line for comparison.}
\label{fig:TC}
\end{center}
\end{figure}

The net charge $q$ localized at the DW is obtained
by integrating the charge distribution over the whole system,
\begin{align}
q &= \sum_z \delta\rho(z)a \\
&= \frac{e}{N}\sum_{k_x,k_y,z}\sum_{\epsilon \leq\mu} \left[ |\psi_{k_x,k_y,n}^{\rm (DW)}(z)|^2 - |\psi_{k_x,k_y,n}^{\rm (uniform)}(z)|^2 \right], \nonumber
\end{align}
per unit area of the DW.
Figure \ref{fig:TC} shows the behavior of the total charge as a function of the chemical potential $\mu$
for each DW configuration,
with the DW width $W$ fixed to $0.1 L_z$.
We can see that the \textit{spiral} DW bears the largest charge
and the \textit{head-to-head} DW the smallest,
as we have seen in the charge distribution calculation.
It should be noted that the total charge is almost proportional to the chemical potential $\mu$ for any DW configuration,
since the number of the occupied states in the lowest LL is proportional to $\mu$
due to its constant density of states around $\mu=0$.
When the chemical potential $\mu$ goes beyond the energy scale $\sim 0.1t$,
which is not shown in Fig.~\ref{fig:TC},
the total charge $q$ becomes no longer proportional to $\mu$,
since not only the zeroth LL but the higher LLs also gets occupied and contribute to the localized charge.

Under a certain DW configuration,
the Weyl equation under the continuum Hamiltonian [Eq.~(\ref{eq:cont-Hamiltonian})] can be analytically solved
to estimate the amount of the localized charge.
In Ref.~\onlinecite{Araki_DW},
the authors employed the simplified \textit{collinear} DW configuration,
which is described by the magnetic texture
\begin{align}
\boldsymbol{M}_\mathrm{coll}(z) = M_0 \left( \tanh\frac{z}{W},\ 0, \ 0 \right),
\end{align}
and derived the net localized charge
\begin{align}
q_\mathrm{coll} &= \frac{e}{\pi^2} \frac{k_{\Delta}}{v_{\rm F}} \mu \label{eq:charge-collinear}
\end{align}
by using the distribution of the analytically obtained wavefunctions in the zeroth LL.
Since the axial magnetic field structure of the \textit{collinear} DW is identical to that of the \textit{coplanar} DW, 
this analytical form applies to the \textit{coplanar} DW as well.
In Fig.~\ref{fig:TC}, we also show this analytical form in the solid red line.
It is in good agreement with the numerically obtained $q$ for the \textit{coplanar} DW,
from which we can ensure that the charge localization around the DW is
dominated by the zeroth LL, as assumed in Ref.~\onlinecite{Araki_DW}.

The localized charge confirmed here implies that,
when the DW is moving adiabatically (e.g. by an external magnetic field),
it carries the localized charge, resulting in a current pulse.
Based on this charge pumping picture,
the localized charge was conversely estimated in Ref.~\onlinecite{Araki_pumping},
as
\begin{align}
q = \frac{e}{2\pi^2} \frac{J \mu}{v_\mathrm{F}^2} \int dz |\partial_z \boldsymbol{M}_\perp(z)|
 = \frac{e^2}{2\pi^2} \frac{\mu}{v_\mathrm{F}} \int dz |\boldsymbol{B}_{5\perp}(z)|, \label{eq:pumping}
\end{align}
where the subscript $\perp$ denotes vector components transverse to the $z$-axis.
This relation was derived by assuming the quantum limit of the Landau quantization,
requiring the Fermi level $\mu$ to be low enough so that it should be between the zeroth and first LLs
under the axial magnetic field $\boldsymbol{B}_5(z)$ throughout the whole system.
Under this assumption, Eq.~(\ref{eq:pumping}) gives the localized charge per unit area
of the \textit{coplanar}(c), \textit{spiral}(s), and \textit{head-to-head}(h) DWs as
\begin{align}
q_\mathrm{c} = q_\mathrm{h} = \frac{e}{\pi^2} \frac{JM_0}{v_\mathrm{F}^2} \mu, \quad q_\mathrm{s} = \frac{e}{2\pi} \frac{JM_0}{v_\mathrm{F}} \mu. \label{eq:pumping-DW}
\end{align}
For the \textit{coplanar} DW, this relation is consistent with that from the analytical diagonalization given in Eq.~(\ref{eq:charge-collinear}),
namely the solid red line in Fig.~\ref{fig:TC}.
Comparing these relations with our numerical results in the present work,
we find that Eq.~(\ref{eq:pumping-DW}) overestimates the localized charge for the \textit{head-to-head} DW.
Such a discrepancy can presumably be traced back to the breakdown of the quantum limit assumption;
since the axial magnetic field drops sharply to zero at the center of the \textit{head-to-head} DW,
the quantum limit assumption is no longer reliable about the center of the DW.
On the other hand, Eq.~(\ref{eq:pumping-DW}) well describes our numerical results
for the \textit{coplanar} and \textit{spiral} DWs,
since the axial magnetic field for these DWs is strong enough to apply the quantum limit assumption
within the region of the DW.
Since the size of the system is finite in our numerical calculation,
we lose the ``tail'' of the localized charge beyond the boundary,
which slightly reduces the net charge $q$ from its analytically obtained value in the infinite system.

\subsection{Physical implications}
Let us now discuss how the localized charge at the DW can be captured experimentally.
Regarding the system as Cr-doped $\mathrm{Bi_2 Se_3}$ (Ref.~\onlinecite{Kurebayashi_2014}),
we can estimate the amount of the localized charge with Eq.~(\ref{eq:charge-collinear})
for the \textit{coplanar} DW.
We take the Fermi velocity (within the quintuple layer) as $v_\mathrm{F} = 3.55\mathrm{eV}\text{\AA}$,
which was observed by the ARPES measurement \cite{Kuroda_2010}.
From the first-principles calculation \cite{Yu_R},
the exchange coupling between the localized magnetic moment of Cr and the \textit{p}-electron
was estimated as $JM_0 = 2.7 \mathrm{eV}$.
In the present system,
since Cr is doped by partially substituting Bi to the concentration $x$,
we use $x JM_0$ instead of $JM_0$ in our calculation,
by taking the local average of the exchange energy around each site;
here we use $x=0.05$, which was suggested by Ref.~\onlinecite{Kurebayashi_2014}.
By using the parameters given above,
and tuning the chemical potential $\mu$ to $100 \mathrm{meV}$,
the localized charge on the \textit{coplanar} DW is obtained as
\begin{align}
q = 1.1 e \times 10^{11} \mathrm{cm}^{-2}. \label{eq:localized-charge}
\end{align}
Therefore, for a DW with its size $W=10\mathrm{nm}$,
the average charge density inside the DW is given as
$q/W = 1.1e \times 10^{18} \mathrm{cm}^{-3}$.
This value is comparable to the typical charge density in GaAs,
which implies that the localized charge may be directly measured by the scanning tunneling microscope (STM)
in a thin-film geometry \cite{Wiesendanger}.

Moreover, since the DW behaves as a charged object,
we can expect that an external electric field applied to this system can drive the motion of the DW.
Here we should note that the electron transmission beyond the DW is suppressed
if the chirality-flipping scattering process is negligible,
since the DW drastically exchange the positions of the valleys in the Brillouin zone \cite{Kobayashi_2018}.
Therefore,
the DW in our magnetic WSM is driven only by the electrostatic force,
which is distinct from the well-known spin-transfer and spin-orbit torques
invoked by the conduction electron.
Such a driving force may be more advantageous than that by the current-induced torques,
in that it can reduce the energy loss by Joule heating.

The creep velocity of the DW can be roughly estimated by the semiclassical equation of motion for a DW.
In the absence of a pinning potential,
the dynamics of the DW obeys a simple Newtonian equation of motion \cite{Doring}
\begin{align}
M_W \ddot{X} + \frac{M_W}{\tau_W}\dot{X} = qE \label{eq:DW-motion}
\end{align}
for a DW per unit area under the electric field $E$,
where $X$ denotes the center of mass of the DW in the collective coordinate representation.
The ``inertial mass'' (per unit area) $M_W = N_W / KW^2$ comes from
the moment of inertia of each localized spin against the external torque,
while the ``friction factor'' $\tau_W^{-1} = \alpha K/\hbar$ arises from the Gilbert damping;
here $N_W = 2xW/a^3$ is the number of localized spins in a unit area of the DW,
$a$ is the lattice constant, $K$ is the magnetic anisotropy energy, and $\alpha$ is the Gilbert damping constant.
If the Gilbert damping is sufficiently large,
the driving force here leads to an adiabatic motion of the DW.
The DW velocity $V$ eventually reaches a constant value, namely the creep velocity,
\begin{align}
V_\mathrm{C} = \frac{\tau_W}{M_W} q E = \frac{a^3 W}{2\alpha x} q E.
\end{align}
We should note that this creep velocity becomes independent of the Cr concentration $x$
as long as the localized charge $q$ comes only from the zeroth LL,
since $q$ is proportional to the (averaged) exchange energy $xJM_0$ in this regime,
as we have seen in the discussion above.
Using the lattice constant $a=9.84\mathrm{\AA}$
employed for the first-principles calculation on $\mathrm{Bi_2 Se_3}$ (Ref.~\onlinecite{Mishra})
and fixing the Gilbert damping constant to the typical value $\alpha =0.01$,
the localized charge $q$ obtained in Eq.~(\ref{eq:localized-charge}) for the \textit{coplanar} DW
yields the creep velocity $V_\mathrm{C} = 81 \mathrm{m/s}$
under the electric field $E = 100 \mathrm{V/cm}$.
This velocity is not as fast as the typical DW velocity driven by spin-orbit torques,
which can reach up to 300-400\,m/s (see Refs.~\onlinecite{Emori} and \onlinecite{Ryu_K-S}).
However, our method using WSMs is still advantageous in the energetic efficiency,
i.e. the reduction of the Joule heating.

\section{Conclusion} 
\label{sec:conclusion}
In this paper, we have investigated the electronic properties of a magnetic WSM in the presence of magnetic DWs.
Using the energy eigenvalues and eigenstates obtained from the numerical diagonalization of the magnetic Weyl Hamiltonian,
we have calculated the DW excitation energy and the charge distribution for three types of DWs:
\textit{coplanar}, \textit{spiral}, and \textit{head-to-head}.
The main finding in this paper is that a DW in a magnetic WSM induces a localized charge at the DW,
the amount of which linearly depends on the Fermi level for any type of DW at low energy.
We can understand this behavior by regarding the magnetic texture as an emergent axial magnetic field for the electrons:
as the magnetic DW texture is equivalent to the axial magnetic field around the center of the DW,
it yields Landau quantized states at the DW,
and the zeroth LL among them contributes to the localized charge at low energy.
Since the zeroth LL in 3D is unidirectionally dispersed and shows constant density of states at low energy,
the localized charge, namely the number of the occupied states, becomes proportional to the Fermi energy.
The charge distribution around the DW qualitatively agrees with the spatial profile of the axial magnetic field for each DW configuration,
which also supports our ``Landau quantization'' hypothesis.

While we have neglected the Coulomb interaction in our analysis,
it usually has a significant effect on the charge distribution:
when an electric charge is put inside a conductor,
a charge with the opposite sign gets accumulated around the original charge due to the Coulomb interaction,
which is known as the electrostatic screening.
Although the localized charge in a WSM should also be subject to the screening in the presence of the Coulomb interaction,
we expect that it would be less significant than that in normal metals.
The screening effect in general can be neglected
if the system size is smaller than the screening length:
this is more likely to occur in WSMs than in normal metals,
since the screening length in WSMs at low energy tends to be longer than that in normal metals
due to the vanishing density of states at the Weyl points.

Making use of this localized charge,
we have semiclassically demonstrated that
a DW in a magnetic WSM can be driven by the electrostatic force from an external electric field.
This effect can be regarded as the inverse effect of the charge pumping conveyed by the dynamics of a magnetic texture in a WSM,
which was proposed in the recent paper by the authors \cite{Araki_pumping}.
It should be noted that the DW in this system almost totally suppresses the conduction current \cite{Kobayashi_2018}
and thus will not suffer from energy loss by Joule heating,
which is quite advantageous in making use of such DWs for spintronics.
Recent transport and ARPES measurements found that a Heusler alloy $\mathrm{Co_3 Sn_2 S_2}$ shows the ferromagnetic WSM phase \cite{Liu_E,Muechler,Xu_Q},
which may serve as a good playground for magnetic DWs treated in our work.
On the other hand, our hypothetical lattice model cannot be straightforwardly applied to
antiferromagnetic WSMs,
such as $\mathrm{Mn_3 Sn}$ (Refs.~\onlinecite{Yang_2017,Ito,Kuroda}),
since noncollinear antiferromagnetic orders on the kagome lattice are expected in those materials.
In antiferromagnetic $\mathrm{Mn_3 Sn}$, for instance,
the tight-binding model calculation \cite{Ito} showed that
its chiral antiferromagnetic order is related to the separation of the Weyl points and the anomalous Hall conductivity.
Therefore, we can make a rough qualitative assumption that
a spatial modulation of the antiferromagnetic order becomes equivalent to the axial magnetic field,
which implies that an antiferromagnetic DW may host the localized charge in a similar manner to the ferromagnetic DWs observed here.

\acknowledgments{
The authors appreciate K.~Kobayashi, D.~Kurebayashi, and Y.~Ominato for fruitful discussions at Tohoku University.
Y.~A. is supported by JSPS KAKENHI Grant Number JP17K14316.
K.~N. is supported by JSPS KAKENHI Grant Numbers JP15H05854 and JP17K05485.
}


\begin{thebibliography}{99}

\bibitem{Wan_2011}
X.~Wan, A.~M.~Turner, A.~Vishwanath, and S.~Y.~Savrasov,
Phys.~Rev.~B \textbf{83}, 205101 (2011).

\bibitem{Burkov_2011}
A.~A.~Burkov and L.~Balents,
Phys.~Rev.~Lett.~\textbf{107}, 127205 (2011).

\bibitem{Zyuzin_2012}
A.~A.~Zyuzin, S.~Wu, and A.~A.~Burkov,
Phys.~Rev.~B \textbf{85}, 165110 (2012).

\bibitem{Young_2012}
S.~M.~Young, S.~Zaheer, J.~C.~Y.~Teo, C.~L.~Kane, E.~J.~Mele, and A.~M.~Rappe,
Phys.~Rev.~Lett.~\textbf{108}, 140405 (2012).

\bibitem{Xu_S-Y_2015}
S.-Y.~Xu, I.~Belopolski, N.~Alidoust, M.~Neupane, G.~Bian, C.~Zhang, R.~Sankar, G.~Chang, Z.~Yuan, C.-C.~Lee, S.-M.~Huang, H.~Zheng, J.~Ma, D.~S.~Sanchez, B.~Wang, A.~Bansil, F.~Chou, P.~P.~Shibayev, H.~, S.~Jia, and M.~Zahid Hasan,
Science \textbf{349}, 613 (2015).

\bibitem{Lv_B-Q_2015}
B.~Q.~Lv, H.~M.~Weng, B.~B.~Fu, X.~P.~Wang, H.~Miao, J.~Ma, P.~Richard, X.~C.~Huang, L.~X.~Zhao, G.~F.~Chen, Z.~Fang, X.~Dai, T.~Qian, and H.~Ding,
Phys.~Rev.~X \textbf{5}, 031013 (2015).

\bibitem{Lv_B-Q_2015_2}
B.~Q.~Lv, N.~Xu, H.~M.~Weng, J.~Z.~Ma, P.~Richard, X.~C.~Huang, L.~X.~Zhao, G.~F.~Chen, C.~Matt, F.~Bisti, V.~Strokov, J.~Mesot, Z.~Fang, X.~Dai, T.~Qian, M.~Shi, and H.~Ding,
Nat.~Phys.~\textbf{11}, 724 (2015).

\bibitem{Xu_S-Y_2015_2}
S.-Y.~Xu, N.~Alidoust, I.~Belopolski, C.~Zhang, G.~Bian, T.-R.~Chang, H.~Zheng, V.~Strokov, D.~S.~Sanchez, G.~Chang, Z.~Yuan, D.~Mou, Y.~Wu, L.~Huang, C.-C.~Lee, S.-M.~Huang, B.~Wang, A.~Bansil, H.-T.~Jeng, T.~Neupert, A.~Kaminski, H.~Lin, S.~Jia, and M.~Zahid Hasan,
Nat.~Phys.~\textbf{11}, 748 (2015).

\bibitem{Liu_E}
E.~Liu, Y.~Sun, L.~M\"{u}echler, A.~Sun, L.~Jiao, J.~Kroder, V.~S\"{u}{\ss}, H.~Borrmann, W.~Wang, W.~Schnelle, S.~Wirth, S.~T.~B.~Goennenwein, and C.~Felser,
arXiv:1712.06722.

\bibitem{Muechler}
L.~Muechler, E.~Liu, Q.~Xu, C.~Felser, and Y.~Sun,
arXiv:1712.08115.

\bibitem{Xu_Q}
Q.~Xu, E.~Liu, W.~Shi, L.~Muechler, C.~Felser, and Y.~Sun,
arXiv:1801.00136.

\bibitem{Yang_2017}
H.~Yang, Y.~Sun, Y.~Zhang, W.-J.~Shi, S.~S.~P.~Parkin, and B.~Yan,
New J.~Phys.~\textbf{19}, 015008 (2017).

\bibitem{Ito}
N.~Ito and K.~Nomura,
J.~Phys.~Soc.~Jpn.~\textbf{86}, 063703 (2017).

\bibitem{Kuroda}
K.~Kuroda, T.~Tomita, M.-T.~Suzuki, C.~Bareille, A.~A. Nugroho, P.~Goswami, M.~Ochi, M.~Ikhlas, M.~Nakayama, S.~Akebi, R.~Noguchi, R.~Ishii, N.~Inami, K.~Ono, H.~Kumigashira, A.~Varykhalov, T.~Muro, T.~Koretsune, R.~Arita, S.~Shin, T.~Kondo, and S.~Nakatsuji,
Nat.~Mater.~\textbf{16}, 1090 (2017).

\bibitem{Kurebayashi_2014}
D.~Kurebayashi and K.~Nomura,
J.~Phys.~Soc.~Jpn.~\textbf{83}, 063709 (2014).

\bibitem{Liu_2013}
C.-X.~Liu, P.~Ye, and X.-L.~Qi,
Phys.~Rev.~B \textbf{87}, 235306 (2013).

\bibitem{Grushin}
A.~G.~Grushin, J.~W.~F.~Venderbos, A.~Vishwanath, and R.~Ilan,
Phys.~Rev.~X \textbf{6}, 041046 (2016).

\bibitem{Pikulin}
D.~I.~Pikulin, A.~Chen, and M.~Franz,
Phys.~Rev.~X \textbf{6}, 041021 (2016).

\bibitem{Araki_pumping}
Y.~Araki and K.~Nomura,
arXiv:1711.03135.

\bibitem{Nomura_2011}
K.~Nomura and N.~Nagaosa,
Phys.~Rev.~Lett.~\textbf{106}, 166802 (2011).

\bibitem{Brataas}
A.~Brataas, A.~D.~Kent, and H.~Ohno,
Nat.~Mater.~\textbf{11}, 372 (2012).

\bibitem{Parkin}
S.~S.~P.~Parkin, M.~Hayashi, and L.~Thomas,
Science \textbf{320}, 190 (2008).

\bibitem{Berger}
L.~Berger,
J.~Appl.~Phys.~\textbf{49}, 2156 (1978);
Phys.~Rev.~B \textbf{33}, 1572 (1986).

\bibitem{Saihi}
E.~Salhi and L.~Berger,
J.~Appl.~Phys.~\textbf{73}, 6405 (1993).

\bibitem{Slonczewski}
J.~C.~Slonczewski,
J.~Magn.~Magn.~Mater.~\textbf{159}, L1 (1996).

\bibitem{Ralph}
D.~C.~Ralph and M.~D.~Stiles,
J.~Magn.~Magn.~Mater.~\textbf{320}, 1190 (2008).

\bibitem{Manchon_2008}
A.~Manchon and S.~Zhang,
Phys.~Rev.~B \textbf{78}, 212405 (2008).

\bibitem{Manchon_2009}
A.~Manchon and S.~Zhang,
Phys.~Rev.~B \textbf{79}, 094422 (2009).

\bibitem{Miron_2010}
I.~M.~Miron, G.~Gaudin, S.~Auffret, B.~Rodmacq, A.~Schuhl, S.~Pizzini, J.~Vogel, and P.~Gambardella,
Nat.~Mater.~\textbf{9}, 230 (2010).

\bibitem{Miron_2011}
I.~M.~Miron, K.~Garello, G.~Gaudin, P.-J.~Zermatten, M.~V.~Costache, S.~Auffret, S.~Bandiera, B.~Rodmacq, A.~Schuhl, and P.~Gambardella,
Nature \textbf{476}, 189 (2011).

\bibitem{Emori}
S.~Emori, U.~Bauer, S.-M.~Ahn, E.~Martinez, and G.~S.~D.~Beach,
Nat.~Mater.~\textbf{12}, 611 (2013).

\bibitem{Ryu_K-S}
K.-S.~Ryu, L.~Thomas, S.-H.~Yang, and S.~Parkin,
Nat.~Nanotech.~\textbf{8}, 527 (2013).

\bibitem{Shiomi_2014}
Y.~Shiomi, K.~Nomura, Y.~Kajiwara, K.~Eto, M.~Novak, K.~Segawa, Y.~Ando, and E.~Saitoh,
Phys.~Rev.~Lett.~\textbf{113}, 196601 (2014).

\bibitem{Araki_DW}
Y.~Araki, A.~Yoshida, and K.~Nomura,
Phys.~Rev.~B \textbf{94}, 115312 (2016).

\bibitem{Wilson}
K.~G.~Wilson,
``Quarks and Strings on a Lattice''
in A.~Zichichi eds.~\textit{New Phenomena in Subnuclear Physics}
(Springer, 1977).

\bibitem{Qi_2008}
X.-L.~Qi, T.~L.~Hughes, and S.-C.~Zhang,
Phys.~Rev.~B \textbf{78}, 195424 (2008).

\bibitem{Araki_corr}
Y.~Araki and K.~Nomura,
Phys.~Rev.~B \textbf{93}, 094438 (2016).

\bibitem{K-Y_Yang}
K.-Y.~Yang, Y.-M.~Lu, and Y.~Ran,
Phys.~Rev.~B \textbf{84}, 075129 (2011).

\bibitem{DW-note}
The surfaces host the Fermi arc states,
which may look degenerate with the DW Fermi arc states, or the zeroth LLs under the axial magnetic field.
In order to rule out such contribution from the surfaces,
we take the periodic boundary condition at $z=\pm L_z/2$ with two opposite \textit{coplanar} DWs at $z = \pm L_z/4$
for this calculation only.
To be precise, we take the magnetic texture
\begin{align}
\tilde{\boldsymbol{M}}_\mathrm{c}(z) = M_0 \left(\tanh\frac{z_+}{W}-\tanh\frac{z_-}{W}, \ 0, \ \sech\frac{z_+}{W}-\sech\frac{z_-}{W} \right),
\end{align}
with $z_\pm \equiv z \mp L_z/4$.

\bibitem{Kuroda_2010}
K.~Kuroda, M.~Arita, K.~Miyamoto, M.~Ye, J.~Jiang, A.~Kimura, E.~E.~Krasovskii, E.~V.~Chulkov, H.~Iwasawa, T.~Okuda, K.~Shimada, Y.~Ueda, H.~Namatame, and M.~Taniguchi,
Phys.~Rev.~Lett.~\textbf{105}, 076802 (2010).

\bibitem{Yu_R}
R.~Yu, W.~Zhang, H.-J.~Zhang, S.-C.~Zhang, X.~Dai, and Z.~Fang,
Science \textbf{329}, 61 (2010).

\bibitem{Wiesendanger}
R.~Wiesendanger,
\textit{Scanning Probe Microscopy and Spectroscopy: Methods and Applications}
(Cambridge University Press, 1994).

\bibitem{Kobayashi_2018}
K.~Kobayashi, Y.~Ominato, and K.~Nomura,
arXiv:1802.04536.

\bibitem{Doring}
W.~D\"{o}ring,
Z.~Naturforsch.~\textbf{3A}, 373 (1948).

\bibitem{Mishra}
S.~K.~Mishra, S.~Satpathy, and O~Jepsen,
J.~Phys.: Condens.~Mattter \textbf{9}, 461 (1997).

\end{thebibliography}
\end{document}